\documentclass[12pt, draftcls, onecolumn]{IEEEtran}

\ifCLASSINFOpdf
  \usepackage[pdftex]{graphicx}
  \DeclareGraphicsExtensions{.pdf,.jpeg,.png}
\else
  \usepackage[dvips]{graphicx}
\fi
\usepackage{epstopdf}
\usepackage[cmex10]{amsmath}
\usepackage{amssymb}
\usepackage{wasysym}
\usepackage{algorithm}
\usepackage{algorithmic}
\usepackage{array}
\usepackage{cite}
\usepackage{color}
\usepackage{url}

\usepackage{epsfig,latexsym}
\usepackage{flushend}
\usepackage{cite}
\usepackage{verbatim}
\usepackage{amsopn}
\usepackage{booktabs}

\begin{document}
%
\title{Precoding Design for Energy Efficiency of Multibeam Satellite Communications}


\author{Chenhao~Qi,~\IEEEmembership{Senior~Member,~IEEE} and Xin~Wang~\IEEEmembership{Student~Member,~IEEE}
\thanks{This work is supported in part by National Natural Science Foundation of China under Grant 61302097 and Natural Science Foundation of Jiangsu Province under Grant BK20161428. (\textit{Corresponding author: Chenhao~Qi})}
\thanks{Chenhao~Qi and Xin~Wang are with the School of Information Science and Engineering, Southeast University, Nanjing 210096, China (Email: qch@seu.edu.cn).} }


\markboth{}
{Shell \MakeLowercase{\textit{et al.}}: Bare Demo of IEEEtran.cls for Journals}
\vspace{-8em}
\maketitle
\vskip 8.5em
\begin{abstract}
Instead of merely improving the spectral efficiency (SE), improving the energy efficiency (EE) is another important concern for multibeam satellite systems, due to the power constraint of satellites. However, so far there has been no detailed work on the precoding design concerning the EE for multibeam satellite. In this work, the EE maximization problem is investigated for multibeam satellite systems under the total power constraint as well as the quality of service (QoS) constraints. Precoding design algorithms based on zero forcing (ZF) and sequential convex approximation (SCA) are presented respectively. In particular, these algorithms are verified by the real measured channel data of multibeam satellite systems. Numerical results show that the precoding algorithm based on SCA outperforms that based on ZF.  It is also implied that the EE cannot be always improved by solely increasing the power of the satellite, while reducing the satellite operation power is an effective way for the EE improvement.

\end{abstract}

\section{Introduction}
With rapid development of satellite manufacturing, multibeam satellite communication is a promising candidate for the next generation satellite communications due to its high spectral efficiency (SE)~\cite{2LowComplex}. To support the terabit capacity, full frequency reuse among beams is attractive since larger bandwidth can be provided for each user~\cite{FutureChallenges,LMSchannelModel}. As a consequence, precoding is required for multibeam satellite systems to mitigate inter-beam interference so as to improve the SE. In~\cite{4}, a generic iterative algorithm for SE maximization with linear power constraints is proposed to optimize the precoding and power allocation alternatively for unicast multibeam satellite systems. Then in~\cite{FrameBased}, multicast multibeam satellite systems is considered, where the precoding and power allocation are jointly optimized under the power constraints of each beam. More recently in~\cite{2LowComplex}, a robust precoding scheme for multicast multibeam satellite system is proposed based on a first perturbation model, considering that the channel state information will be corrupted at the satellite gateway.

The aforementioned works only consider SE of multibeam satellite systems, while the total power consumption is not taken into account. Note that the satellite is usually powered by solar battery. The power consumption of the satellite is nonnegligible. Energy efficiency (EE), defined as the ratio of the system throughput over total power consumption, is an important factor for multibeam satellite systems. In fact, EE maximization has already been extensively studied in terrestrial wireless communications~\cite{MainUse,QiChenhao}. Inspired by these work, we consider the EE aspect for multibeam satellite system. Improving EE can reduce the satellite size and extend the satellite lifetime. The power amplifier of the transponder can operate linearly, avoiding non-linearity and intermodulation products. Currently, the work on EE maximization for multibeam satellite systems is only reported by~\cite{6}. However, the detailed steps for precoding are not clear and the constraints of quality of service (QoS) for different users are not considered.

In this letter, we consider the EE maximization problem for multibeam satellite systems under the total power constraint and the QoS constraints. We present two precoding algorithms based on zero forcing (ZF) and sequential convex approximation (SCA). In the first algorithm, we use the Dinkelbach's method to solve the fractional programming. In the second algorithm, we sequentially convert the original nonconvex problem by SCA and finally approximate it as a convex optimization problem. The detailed steps for the precoding design are provided. In particular, the algorithms are verified by the measured channel data of multibeam satellite systems.

The notations are defined as follows. $\mathcal{U}$, $\mathcal{CN}$, $\mathbb{R}$ and $\mathbb{C}$ represent the uniform distribution, complex Gaussian distribution, set of real numbers and set of complex numbers. $x^{(n)}$ represents the value of $x$ after the $n$th iteration.

\section{System Model}\label{sec.SystemModel}
We consider a broadband satellite system which provides service to fixed users via multiple beams. The array feed reflector transforms $N$ feed signals into $K$ transmitted signals. By using time division multiplexing (TDM), a single user per beam is scheduled at each time slot. To improve the spectral efficiency, full frequency reuse is considered.

Based on the above settings, the multibeam satellite channel $\boldsymbol{H}\in\mathbb{C}^{K\times N}$ from the satellite to users can be modeled as \cite{FrameBased}
\begin{equation}
\boldsymbol{H} = \boldsymbol{\varPhi}\boldsymbol{A},
\end{equation}
where $\boldsymbol{\varPhi}\in\mathbb{C}^{K\times K}$ represents the phase variation effects due to different propagation paths among the satellite and the users, and $\boldsymbol{A}\in\mathbb{R}^{K\times N}$ represents the multibeam antenna pattern.

Since the satellite antenna feed spacing is relatively small compared to the long propagation path, the phases among one user and all antenna feeds are commonly assumed to be identical in line-of-sight (LOS) environment \cite{4}. Hence, $\boldsymbol{\varPhi}$ is a diagonal matrix with the $i$th diagonal entry defined as $[\boldsymbol{\varPhi}]_{i,i} \triangleq e^{j\phi_i},i=1,...,K$, where $\phi_i$ denotes a uniformly distributed variable, i.e., $\phi_i \sim \mathcal{U}(0,2\pi)$.

The entry at the $k$th row and $n$th column of $\boldsymbol{A}$ is given by
\begin{equation}\label{Equal.Channel}
a_{k,n} = \frac{\sqrt{G_R G_{k,n}}}{4 \pi \frac{d_k}{\lambda}\sqrt{\kappa T_R B_W}},
\end{equation}
where $G_R$, $G_{k,n}$ and $d_k$ denote the receiving antenna gain of the users, the gain between the $n$th feed and the $k$th user, and the distance between the satellite and the $k$th user, respectively. $\lambda$, $B_W$, $\kappa$ and $T_R$ are the wavelength, the bandwidth, Boltzmann constant, and the clear sky noise temperature of the receiver, respectively.

The received signal is
\begin{equation}
\boldsymbol{y} = \boldsymbol{H}\boldsymbol{W}\boldsymbol{x} + \boldsymbol{n},
\end{equation}
where $\boldsymbol{y}\in\mathbb{C}^{K\times 1}$ is a signal vector received by $K$ users, $\boldsymbol{W}\in\mathbb{C}^{N\times K}$ is a precoding matrix to be designed, $\boldsymbol{x}\in\mathbb{C}^{K\times 1}$ is the data to be transmitted to the users, and $\boldsymbol{n}\in\mathbb{C}^{K\times 1}$ is an additive white Gaussian noise (AWGN) vector with each entry identically and independently distributed, i.e., $\boldsymbol{n} \sim \mathcal{CN}(0,\sigma^2 \boldsymbol{I}_K)$.  We further define  $\boldsymbol{H} \triangleq [\boldsymbol{h}_1^T,\boldsymbol{h}_2^T,...,\boldsymbol{h}_K^T]^T$ and $\boldsymbol{W} \triangleq [\boldsymbol{w}_1,\boldsymbol{w}_2,...,\boldsymbol{w}_K]$, where $\boldsymbol{h}_k\in\mathbb{C}^{1\times N}$ is the channel row vector from the satellite to the $k$th user and $\boldsymbol{w}_k\in\mathbb{C}^{N\times 1}$ is the $k$th column of $\boldsymbol{W}$. Therefore, the received signal of the $k$th user can be written as
\begin{equation}
y_k = \boldsymbol{h}_k \boldsymbol{w}_k x_k + \sum_{j \in \mathcal{K},j\neq k}
\boldsymbol{h}_k \boldsymbol{w}_j x_j + n_k,\ k \in \mathcal{K},
\end{equation}
where $x_k$ is the $k$th entry of $\boldsymbol{x}$ representing the data intended for the $k$th user, $n_k$ is the $k$th entry of $\boldsymbol{n}$, and $\mathcal{K}\triangleq  \left\lbrace 1,2,...,K\right\rbrace$ is an user set. For simplicity, we assume the power of the data symbols is normalized, i.e., $|x_k|=1, k\in \mathcal{K}$.

The signal-to-interference-and-noise ratio (SINR) of the $k$th user is
\begin{equation}\label{Equal.SINR}
\mit\Gamma_k = \frac{{|\boldsymbol{h}_k \boldsymbol{w}_k}|^2} {\sum_{j \in \mathcal{K},j\neq k}
	\left| \boldsymbol{h}_k \boldsymbol{w}_j \right| ^2 + \sigma^2},\ k \in \mathcal{K}.
\end{equation}

The total power consumed by the platform and the payloads of satellite is supplied by the solar wings and battery. The platform power consumption is on the same order of magnitude as the payloads power consumption. Generally, the payloads power consumption mainly includes the power consumed by the power amplifiers for the user link, the feeder link and the remote sensing and control link, as well as the on-board signal units. Since different satellite has different parameters, we denote the power consumption of satellite platform generally as $P_0 $.
Now we can formulate the problem of energy efficiency maximization as
\begin{subequations}\label{Equal.Original}
	\begin{align}
	\max_{\boldsymbol{W}} &\ \frac{ B_W\sum_{k \in \mathcal{K}} \ln(1+\mit\Gamma_k)} {\sum_{k \in \mathcal{K}}\| \boldsymbol{w}_k\|_2^2 +P_0} \\
	\textrm{s.t.}  & \quad  \sum_{k\in\mathcal{K}} \| \boldsymbol{w}_k\|_2^2 \leq P_T, \\
	& \quad  \mit\Gamma_k \geq \bar{\mit\Gamma}_k,~  k \in \mathcal{K},
	\end{align}
\end{subequations}
where $P_T$ is the maximum transmission power defined by the power amplifier on the satellite, $\bar{\mit\Gamma}_k$ is the threshold related to the QoS constraint of the $k$th user. Since $B_W$ is a constant, we drop it in the deviarion of the algorithms in order to ease the notations.

\section{Energy efficient Precoding}\label{sec.Algorithms}
\subsection{ZF-based Precoding}\label{subsec.ZF}
By using ZF precoding, where the interference among the users can be entirely eliminated, we define $
\boldsymbol{B} \triangleq \boldsymbol{H}^{H} ( \boldsymbol{H} \boldsymbol{H}^H)^{-1}$ and denote $\boldsymbol{b}_k$ as the $k$th column of $\boldsymbol{B}$. We have
\begin{equation}\label{ObtainWk}
\boldsymbol{w}_k = p_k \frac{\boldsymbol{b}_k}{\|\boldsymbol{b}_k\|_2^2},\  k \in \mathcal{K},
\end{equation}
where $p_k$ is the power of $\boldsymbol{w}_k$. Therefore, (\ref{Equal.SINR}) is rewritten as
\begin{equation}
\mit\Gamma_k = \frac{|\boldsymbol{h}_k \boldsymbol{w}_k|^2}{\sigma^2}
= \frac{|p_k|^2 c_k}{\sigma^2},\  k \in \mathcal{K},
\end{equation}
where $c_k \triangleq |\boldsymbol{h}_k  \boldsymbol{b}_k|^2 / \|\boldsymbol{b}_k\|_2^4$ is solely determined by $\boldsymbol{H}$. We further define $\alpha_k \triangleq |p_k|^2, k \in \mathcal{K}$. Then the design of $\boldsymbol{W}$ in (\ref{Equal.Original}) is converted to the design of $\alpha_k, k \in \mathcal{K}$, where (\ref{Equal.Original}) can be written as
\begin{subequations}\label{Problem9}
	\begin{align}
	\max_{\alpha_k} &\ \frac{\sum_{k\in \mathcal{K}} \ln\left( 1+ \alpha_k c_k/\sigma^2\right) }
	{\sum_{k\in \mathcal{K}} \alpha_k + P_0}\\
	\textrm{s.t.}  &\quad \sum_{k\in\mathcal{K}} \alpha_k \leq P_T,\\
	&\alpha_k \geq \sigma^2 \bar{\mit\Gamma}_k / c_k ,\ k \in \mathcal{K}.
	\end{align}
\end{subequations}
Using Dinkelbach’s method and introducing a Lagrange multiplier $\lambda$, the new optimization problem can be
expressed as
\begin{subequations}
	\begin{align}
	\max_{\alpha_k} &\
	\sum_{k\in \mathcal{K}} \ln{\left( 1+ \frac{\alpha_k c_k}{\sigma^2}\right)  - \Big(\mu+\lambda\Big)   \sum_{k\in\mathcal{K}} \alpha_k }    \label{Equal.Main} \\
	\textrm{s.t.}  &\quad \alpha_k \geq \sigma^2 \bar{\mit\Gamma}_k / c_k, \  k \in \mathcal{K},\label{Equal.Const}
	\end{align}
\end{subequations}
where the constant $\lambda P_T - \mu P_0$ in the objective function is ignored. Note that (\ref{Equal.Main}) can be divided into $K$ independent subproblems with respect to $\alpha_k$, where the $k$th subproblem can be expressed as
\begin{subequations}\label{Equal.subproblem}
	\begin{align}
	 \max_{\alpha_k }&~ \ln
	\left( 1 + \frac{\alpha_k c_k}{\sigma^2}\right) - (\mu + \lambda)\alpha_k\triangleq L_k \\
	\textrm{s.t.} &\quad \alpha_k \geq \frac{\sigma^2 \bar{\mit\Gamma}_k}{c_k},
	\  k \in \mathcal{K}.
	\end{align}
\end{subequations}
Let $\frac{\partial L_k}{\partial \alpha_k}=0$, we can obtain $\alpha_k=\left( \frac{1}{\mu + \lambda}-\frac{\sigma^2}{c_k}\right) $. Therefore, the optimal solution of (\ref{Equal.subproblem}) is
\begin{equation}
\alpha_k^* = \max\left\lbrace \left( \frac{1}{\mu + \lambda} - \frac{\sigma^2}{c_k}\right) , \frac{\sigma^2 \bar{\mit\Gamma}_k}{c_k}\right\rbrace  \label{Equal.Choice}
\end{equation}
for given $\mu$ and $\lambda$. In fact, $\lambda$ can be obtained via bisection search while $\mu$ can be determined by iterative algorithms.

The proposed energy efficient precoding algorithm based on ZF is presented in Algorithm \ref{Algo.ZF-based}. First, we initialize $\mu$ to be zero, i.e., $\mu^{(0)}=0$. The lower bound and upper bound for the bisection search are initialized to be $\lambda_L=0$ and $\lambda_U=1000$, respectively. $\epsilon$ and $\xi$ are used to control the stop condition of the bisection search and Algorithm \ref{Algo.ZF-based}, respectively. The bisection search is included in the steps from step~4 to step~12, where the finally obtained $\alpha_k$ is denoted as $\alpha_k^{(i)}$. Then we obtain $\mu^{(i+1)}$ by
\begin{equation}\label{UpdateW}
  \mu^{(i+1)} = \frac{\sum_{k\in \mathcal{K}} \ln\left( 1+ \alpha_k^{(i)} c_k/\sigma^2\right) }{\sum_{k\in \mathcal{K}} \alpha_k^{(i)} + P_0}.
\end{equation}
We repeat the above procedures until the stop condition is satisfied. Finally we output $\alpha_k^*$, which is the optimized result of $\alpha_k$ through Algorithm~\ref{Algo.ZF-based}. Considering that the phase rotation does not affect the power, we assume $p_k$ is real. Therefore, the designed $\boldsymbol{w}_k$ can be obtained via (\ref{ObtainWk}), where $p_k=\sqrt{\alpha_k^*}$.


\subsection{SCA-based Precoding}
By approximating the EE maximization problem in (\ref{Equal.Original}) as a convex optimization problem, we present a precoding algorithm based on SCA in this section.

We first introduce two variables $t$ and $z$, so that we can rewrite (\ref{Equal.Original}) as
\begin{subequations}\label{Equal.SCA.Orig}
	\begin{align}
	\max_{t,z, \boldsymbol{W}} & \quad \sqrt{t}\\
    \textrm{s.t.} & \ \sum_{k\in \mathcal{K}} \ln\left( 1 + \mit\Gamma_k\right)  \geq \sqrt{tz}, \label{Equal.fenzi} \\
	& \sqrt{z} \geq \sum_{k \in \mathcal{K}} \|\boldsymbol{w}_k\|_2^2 + P_0, \label{Equal.fenmu}\\
	& \sum_{k \in \mathcal{K}} \|\boldsymbol{w}_k\|_2^2 \leq P_T, \label{Equal.TotalPower}\\
	& \mit\Gamma_k \geq \bar{\mit\Gamma}_k, \ k \in \mathcal{K}, \label{Equal.QoS}
	\end{align}
\end{subequations}
where $t$ and $z$ represent squared energy efficiency and squared total power consumption, respectively.


\begin{algorithm}[!t]
	\caption{ZF-based precoding algorithm}
	\label{Algo.ZF-based}
	\begin{algorithmic}[1]
        \STATE \emph{Input:} $\sigma^{2}$, $\bar{\mit\Gamma}_k$, $c_k$, $P_T$.
		\STATE \emph{Initialization:} $i \gets 0$, $\mu^{(0)} \gets 0$, $\lambda_L \gets 0$, $\lambda_U \gets 1000$, $\epsilon \gets 0.1$, $\xi \gets 10^{-3}$.
		\REPEAT
		
		\REPEAT
		\STATE $\lambda \gets (\lambda_L +　\lambda_U)/{2}$.
		\STATE Obtain $\alpha_k,\ k\in \mathcal{K} $ via (\ref{Equal.Choice}).
		\IF {$\sum_{k\in\mathcal{K}}\alpha_k < P_T$}
		\STATE $\lambda_U \gets (\lambda_L +　\lambda_U)/{2}$,
		\ELSE
		\STATE $\lambda_L \gets (\lambda_L +　\lambda_U)/{2}$.
		\ENDIF
		\UNTIL {$|\lambda_U - \lambda_L| \leq \epsilon\  \textrm{and}\  \sum_{k\in \mathcal{K}}\alpha_k \leq P_T$}, where the finally obtained $\alpha_k$ is denoted as $\alpha_k^{(i)}$.
		\STATE Obtain $\mu^{(i+1)}$ via (\ref{UpdateW}). $i=i+1$.
		\UNTIL {$|\mu^{(i)} - \mu^{(i-1)}| \leq \xi$}
		
		\STATE \emph{Output:} $\alpha_k^*$.
	\end{algorithmic}
\end{algorithm}

Based on the fact that the hyperbolic constraint $xy\geq z^2,x\geq 0, y\geq 0$ is equivalent to $\lVert[2z, (x-y)]^T\rVert_2 \leq (x+y)$, (\ref{Equal.fenmu}) can be rewritten in second-order cone (SOC) representation with a newly introduced variable $z'$ as
\begin{equation}
\left\{
\begin{array}{ll}
\frac{z+1}{2} \geq \left\|  \left[ \frac{z-1}{2}, z'\right] ^T\right\| _2\\
\frac{(z' - P_0) + 1}{2} \geq \left\| \left[ \frac{(z'-P_0)-1}{2}, \boldsymbol{w}_1^T, ..., \boldsymbol{w}_K^T\right] ^T\right\| _2.
\end{array}
\right. \label{Equal.fenziTrans}
\end{equation}
Considering the phase rotation does not affect the power, we rewrite (\ref{Equal.QoS}) equivalently as the following SOC representation
\begin{equation}
\left\{
\begin{array}{ll}
\frac{1}{\sqrt{\bar{\mit\Gamma}_k}}  \boldsymbol{h}_k \boldsymbol{w}_k \geq \left( \sigma^2
+ \sum_{j\in \mathcal{K}, j\neq k}|\boldsymbol{h}_k \boldsymbol{w}_j|^2\right)^2\\
\textrm{Im}(\boldsymbol{h}_k \boldsymbol{w}_k) = 0
\end{array}
\right. .\label{Equal.QoS.Trans}
\end{equation}

Since the constraint (\ref{Equal.fenzi}) is still nonconvex, we rewrite (\ref{Equal.fenzi}) as the following two constraints by introducing $\boldsymbol{\gamma}\triangleq [\gamma_1,\gamma_2,...，\gamma_K]^T$ as
\begin{equation}
\sum_{k\in \mathcal{K}} \ln \gamma_k \geq \sqrt{tz}, \label{Equal.Nonconvex1}
\end{equation}
\begin{equation}
1 + \mit\Gamma_k \geq \gamma_k, \ k \in \mathcal{K}. \label{Equal.Nonconvex2}
\end{equation}
Then (\ref{Equal.Nonconvex1}) can be recast as the following two constraints by introducing  $\boldsymbol{\rho}\triangleq [\rho_1,\rho_2,...，\rho_K]^T$ as
\begin{equation}\label{Equal.Rho_t_z}
\sum_{k \in \mathcal{K}} \rho_k \geq \sqrt{tz},
\end{equation}
\begin{equation}
\ln \gamma_k \geq \rho_k \Leftrightarrow \gamma_k \geq e^{\rho_k}, \ k\in \mathcal{K}. \label{Equal.exp}
\end{equation}
It is observed that (\ref{Equal.exp}) is a convex constraint. But $\sqrt{tz}$ in (\ref{Equal.Rho_t_z}) is jointly concave with respect to $t$ and $z$ on the domain $t\geq 0,z\geq 0$. According to \cite{convex}, the convex upper bound is
\begin{align}
\sqrt{tz} \leq \Xi^{(i)}, & \\ \Xi^{(i)}  \triangleq \sqrt{t^{(i)} z^{(i)}} +\frac{t-t^{(i)}}{2} \sqrt{ \frac{z^{(i)}}{t^{(i)}} }
&+ \frac{z-z^{(i)}}{2} \sqrt{ \frac{t^{(i)}}{z^{(i)}} }. \notag\label{Equal.tzUpper}
\end{align}
In fact, $\Xi^{(i)}$ are the first order Taylor series of $\sqrt {tz}$ on the point of $\left(t^{(i)},z^{(i)} \right) $. Then (\ref{Equal.Rho_t_z}) is converted to a linear constraint.


By introducing $\boldsymbol{\beta}\triangleq [\beta_1,\beta_2,...，\beta_K]^T$, we can further rewrite (\ref{Equal.Nonconvex2}) as
\begin{equation}\label{Equal.gamma_beta}
\boldsymbol{h}_k \boldsymbol{w}_k \geq \sqrt{(\gamma_k-1) \beta_k}, \ k\in \mathcal{K},
\end{equation}
\begin{equation}
\beta_k \geq \sigma^2 + \sum_{\begin{subarray}
	\\j \in \mathcal{K} \\j\neq k
	\end{subarray}} |\boldsymbol{h}_k \boldsymbol{w}_j|^2, \ k\in \mathcal{K}. \label{Equal.Beta}
\end{equation}
It is observed that (\ref{Equal.Beta}) is a convex constraint.
\begin{algorithm}[!t]
	\caption{SCA-based precoding algorithm}
	\label{Algo.SCA-based}
	\begin{algorithmic}[1]
		\STATE \emph{Initialization:} $i \gets 0$, $\xi \gets 10^{-3}$.
		\STATE Find any precoding matrix $\boldsymbol{W}^{(0)}$ that satisfies (\ref{Equal.TotalPower}) and (\ref{Equal.QoS.Trans}) as initial value of $\boldsymbol{W}$.
		\STATE Obtain $\mit\Gamma_k^{(0)}$ via (\ref{Equal.SINR}) and then compute $\gamma_k^{(0)}$ via (\ref{Equal.Nonconvex2}).
		\STATE Obtain $\beta_k^{(0)}$ and $z^{(0)}$ via (\ref{Equal.Beta}) and (\ref{Equal.fenmu}), respectively.	
		\STATE Obtain $t^{(0)}$ by $t^{(0)} \gets \left(\sum_{k \in \mathcal{K}}\ln \boldsymbol{\gamma}_k^{(0)} \right)^2 / z^{(0)}$.
		\REPEAT
		\STATE Solve convex optimization problem (\ref{Equal.SCA.Last}) given $\boldsymbol{W}^{(i)}$, $\boldsymbol{\gamma}^{(i)}$, $\boldsymbol{\beta}^{(i)}$, $z^{(i)}$, $t^{(i)}$, where the solutions are denoted as $\boldsymbol{W}^{*}$, $\boldsymbol{\gamma}^{*}$, $\boldsymbol{\beta}^{*}$,$z^{(*)}$, $t^{(*)}$.
		\STATE Update $\boldsymbol{W}^{(i+1)} \gets \boldsymbol{W}^{*}$, $\boldsymbol{\gamma}^{(i+1)} \gets \boldsymbol{\gamma}^*$,
		$\boldsymbol{\beta}^{(i+1)} \gets \boldsymbol{\beta}^{*}$, $z^{(i+1)} \gets z^*$, $t^{(i+1)} \gets t^*$, $i \gets i+1$.
		\UNTIL $|t^{(i)} - t^{(i-1)}| \leq \xi$.
		\STATE \emph{Output:} $\boldsymbol{W}^{*}$.
	\end{algorithmic}
\end{algorithm}


Similarly, the approximation of (\ref{Equal.Rho_t_z}) can also be applied to (\ref{Equal.gamma_beta}) and the convex upper bound of $\sqrt{(\gamma_k -1 )\beta_k}$ is denoted as $\Upsilon_k^{(i)},\ k\in \mathcal{K}$ where
\begin{equation}
\sqrt{(\gamma_k-1)\beta_k} \leq  \Upsilon_k^{(i)},\ k \in \mathcal{K}
\end{equation}
\begin{align}
\Upsilon_k^{(i)} \triangleq & \sqrt{\left( \gamma_k^{(i)}-1\right) \beta_k^{(i)}} +\frac{\gamma_k - \gamma_k^{(i)}}{2} \sqrt{ \beta_k^{(i)} \big/ ( \gamma_k^{(i)}-1 )}   \notag\\
&+\frac{\beta_k - \beta_k^{(i)}}{2} \sqrt{ ( \gamma_k^{(i)}-1 ) \big/ \beta_k^{(i)}} .
\end{align}

Note that maximizing $\sqrt t$ is equivalent as maximizing $t$. (\ref{Equal.SCA.Orig}) is finally converted to a convex optimization problem as
\begin{subequations}\label{Equal.SCA.Last}
	\begin{align}
	\max_{t,z,\boldsymbol{W},\boldsymbol{\gamma},\boldsymbol{\rho},\boldsymbol{\beta}} & \quad t\\
	\textrm{s.t.} \quad & \sum_{k\in \mathcal{K}}\rho_k \geq \Xi^{(i)},\\
	&\boldsymbol{h}_k \boldsymbol{w}_k \geq \Upsilon_k^{(i)},\ k\in \mathcal{K},\\
	&\textrm{(\ref{Equal.TotalPower}), (\ref{Equal.fenziTrans}), (\ref{Equal.QoS.Trans}), (\ref{Equal.exp}), (\ref{Equal.Beta})}.
	\end{align}
\end{subequations}

The proposed energy efficient precoding algorithm based on SCA is outlined in Algorithm~\ref{Algo.SCA-based}. First, we initialize $\boldsymbol{W}$, $\boldsymbol{\gamma}$, $\boldsymbol{\beta}$, $z$, $t$ from step~2 to step~5 as the input of subsequent iterations. Then from step~6 to step~9, we repeat the procedures of solving (\ref{Equal.SCA.Last}) with the CVX\cite{convex} tool and updating the parameters until the stop condition is satisfied. Finally, we output $\boldsymbol{W}^{*}$ as the designed precoding matrix.

\begin{figure}[!t] 
	\centering
	\includegraphics[width=0.82\columnwidth,keepaspectratio]{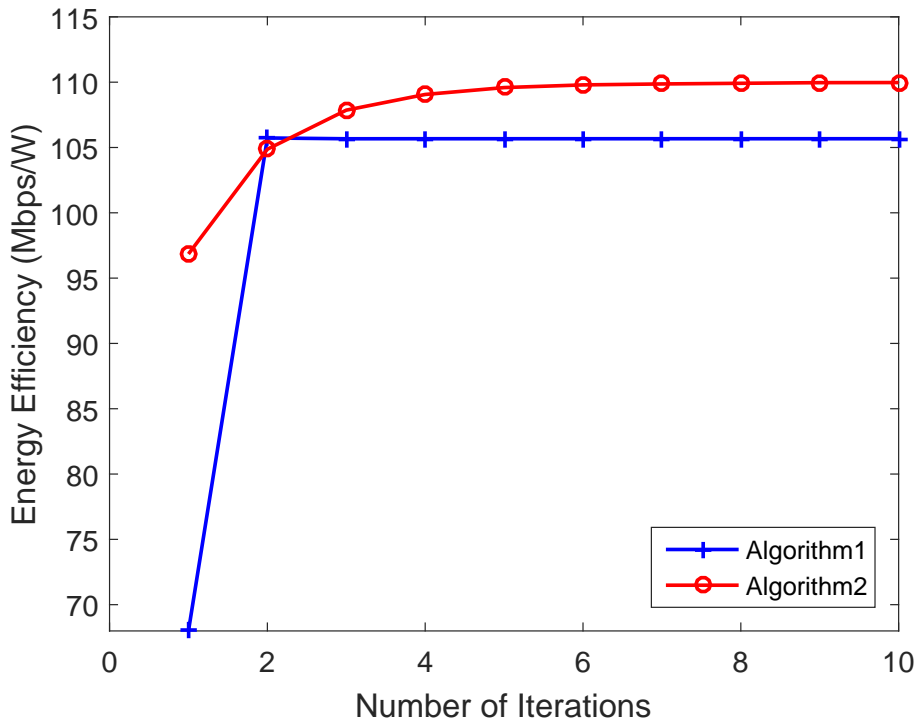}\\
	\caption{Convergence of Algorithm~\ref{Algo.ZF-based} and Algorithm~\ref{Algo.SCA-based}.}
	\label{Fig.Convergence}
\end{figure}

\begin{figure}[!t]
	\centering
	\includegraphics[width=0.82\columnwidth,keepaspectratio]{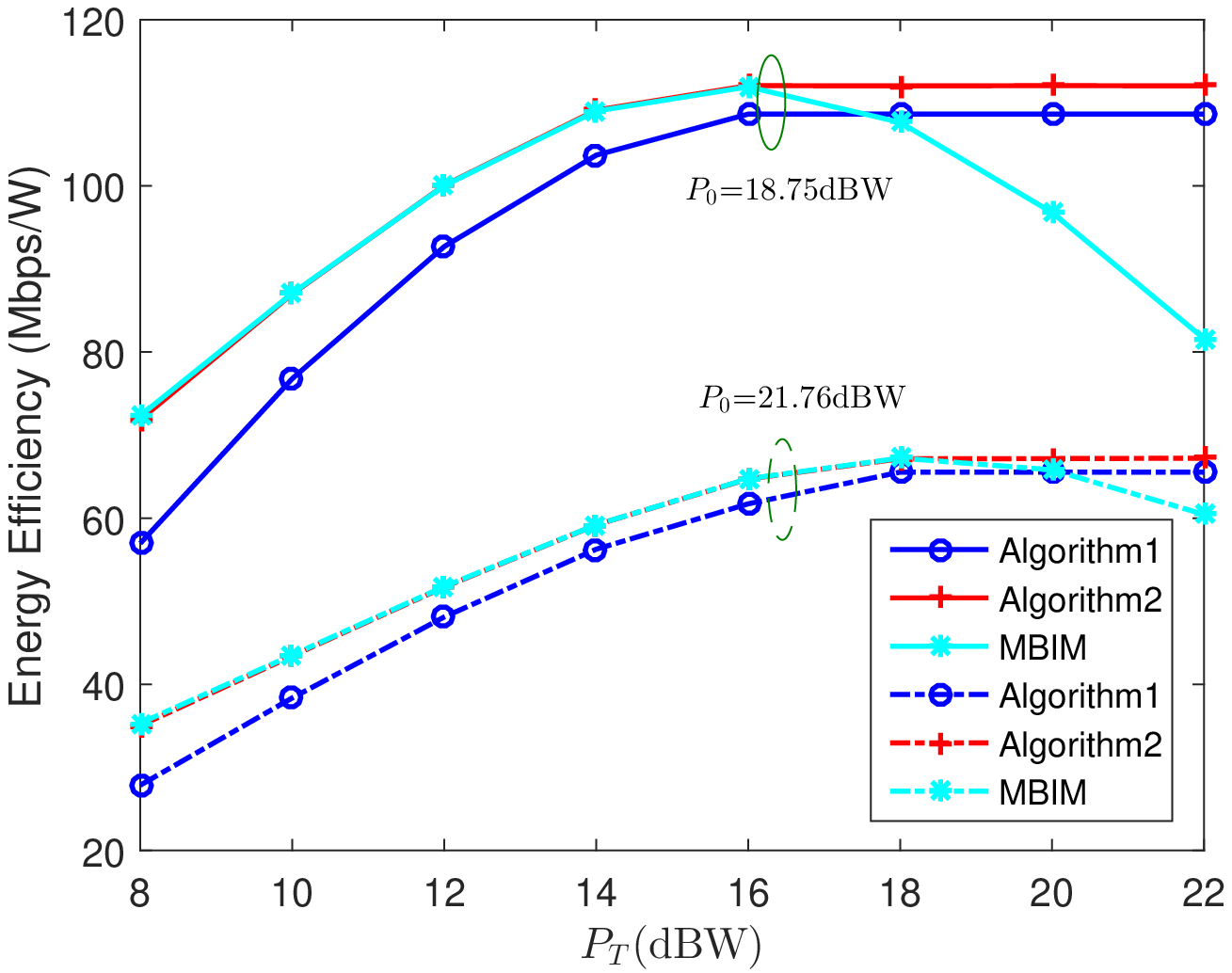}
	\caption{EE comparisons for different $P_T$ and $P_0$.}
	\label{Fig.ConstantPower}
\end{figure}


\section{Numerical results}\label{sec.Simulations}
Now we provide numerical results based on the measured channel data of multibeam satellite systems, which is provided by the European Space Agency (ESA). The multibeam satellite works in the 20GHz Ka band. The user bandwidth, the user antenna gain and  $G/T$ are $500\textrm{MHz}$, $41.7\textrm{dBi}$ and $17.68\textrm{dB/K}$ , respectively. For simplicity, we only consider 7 beams of totally 245 beams that cover the Europe, i.e., $N=K=7$. The SINR thresholds of all users are randomly generated between $-2.85\sim2$dB\cite{FrameBased}. The Boltzmann constant is $1.38\times10^{-23}\textrm{J/K}$ . Since we normalize the noise power by $\kappa T_R B_W$ in (\ref{Equal.Channel}), we set $\sigma^2=1$~\cite{2LowComplex}. The parameter for the stop condition of Algorithm~\ref{Algo.ZF-based} and Algorithm~\ref{Algo.SCA-based} is set to be $\xi=10^{-3}$.

As shown in Fig.~\ref{Fig.Convergence}, we compare the convergence for Algorithm~\ref{Algo.ZF-based} and Algorithm~\ref{Algo.SCA-based}, where we set $P_T=14\textrm{dBW}$ and $P_0=18.75\textrm{dBW}$. It is seen that both Algorithm~\ref{Algo.ZF-based} and Algorithm~\ref{Algo.SCA-based} can fast converge within a small number of iterations. As shown in Fig.~\ref{Fig.ConstantPower}, we compare the EE for different $P_T$ and $P_0$. The multibeam interference mitigation (MBIM) algorithm~\cite{2LowComplex} is also included for comparisons. As $P_T$ increases, three curves first grow, and then the curves of Algorithm~\ref{Algo.ZF-based} and Algorithm~\ref{Algo.SCA-based} get flat while the curve of MBIM falls. The reason that MBIM falls is that the increment of power consumption is faster than that of data rate. It is implied that the EE cannot be always improved by solely increasing the power of the satellite. Therefore, it is not a necessity to equip the satellite with large power for the signal transmission to the users. On the other hand, we also reduce $P_0$ from $P_0=21.76\textrm{dBW}$ to $P_0=18.75\textrm{dBW}$ and make the same simulation. It is seen that around 66.67\% improvement of EE can be achieved with the 13.83\% reduction of $P_0$ for Algorithm~2, which indicating that reducing the constant power consumption is another effective way to improve the EE.



\section{Conclusions}
We have studied the EE maximization problem for multibeam satellite systems under the total power constraint and the QoS constraints. This work may provide a reference to the practical design of multibeam satellite. Future work will focus on the precoding design regarding the tradeoff between the SE and EE.


\bibliographystyle{IEEEtran}
\bibliography{IEEEabrv,IEEEexample}

\end{document}